\documentclass{PoS}

\usepackage{pdflscape}
\usepackage{afterpage}
\usepackage{caption}
\usepackage[flushleft]{threeparttable}
\usepackage{multirow}
\usepackage{url}

\title{Accrete, Accrete, Accrete\ldots Bang! (and repeat):\\The Remarkable Recurrent Novae}

\ShortTitle{The Remarkable Recurrent Novae}

\author{\speaker{Matthew J.\ Darnley}\\
        Astrophysics Research Institute, Liverpool John Moores University, IC2 Liverpool Science Park, 146 Brownlow Hill, Liverpool, L3 5RF, UK\\
        E-mail: \email{M.J.Darnley@ljmu.ac.uk}}

\abstract{All novae recur, but only a handful have been observed in eruption more than once. These systems, the recurrent novae (RNe), are among the most extreme examples of novae. RNe have long been thought of as potential type Ia supernova progenitors, and their claim to this `accolade' has recently been strengthened. In this short review RNe will be presented within the framework of the maximum magnitude---rate of decline (MMRD) phase-space. Recent work integrating He-flashes into nova models, and the subsequent growth of the white dwarf, will be explored. This review also presents an overview of the Galactic and extragalactic populations of RNe, including the newly identified `rapid recurrent nova' subset --- those with recurrence periods of ten years, or less. The most exciting nova system yet discovered --- M31N\,2008-12a, with its annual eruptions and vast nova super-remnant, is introduced. Throughout, open questions regarding RNe, and some of the expected challenges and opportunities that the near future will bring are addressed.}

\FullConference{The Golden Age of Cataclysmic Variables and Related Objects V (GOLDEN2019)\\
		2-7 September 2019\\
		Palermo, Italy}

\begin{document}

\section{Introduction}

Nova eruptions rank among the most luminous stellar {\it explosions} yet discovered. They are surpassed in brightness by only $\gamma$-rays bursts and the full gamut of supernovae (SNe; see, for e.g., \cite{2011PhDT........35K}). Yet novae are substantially more numerous, with $50^{+31}_{-23}$ eruptions every year in the Milky Way \cite{2017ApJ...834..196S}, $65^{+16}_{-15}\,\mathrm{yr}^{-1}$ in the Andromeda Galaxy (M\,31) \cite{2006MNRAS.369..257D}, and as many as 300 annually in M\,87 \cite{2016ApJS..227....1S,2017RNAAS...1a..11S}; compared to $\sim$1\,SN per century per host. Like all explosive transients, novae enrich the ISM; in this case through the production of $^7$Li, $^{13}$C, $^{15}$N, and $^{17}$O (e.g., \cite{2015Natur.518..381T,2018MNRAS.478.1601I,2019arXiv191000575S,Jos16}).

Novae are binary systems. At their heart is a white dwarf (WD) that accretes material from a nearby companion star --- the donor. For main sequence or sub-giant donors, mass-loss occurs via Roche-lobe overflow, and the system is considered a cataclysmic variable (CV). If the donor is a red giant, mass-loss is generally driven by that star's stellar winds, and the system is classed as a symbiotic variable \cite{2012ApJ...746...61D}. In a typical system, material lost from the donor accumulates in an accretion disk around the WD before it is transported to the surface of the later. However, in the presence of a strong WD magnetic field accretion can be partially (intermediate polar) or entirely (polar) channeled via that magnetic field. The reader is referred to \cite{2008clno.book.....B,2014ASPC..490.....W} for compendia of recent detailed and comprehensive reviews of the nova phenomenon. 

Once a critical quantity of material has accreted on to the WD surface, the conditions for nuclear burning are met, which ultimately leads to a thermonuclear runaway under the degenerate conditions of the WD \cite{Jos16,1972ApJ...176..169S,2016PASP..128e1001S}. Once the temperature has increased sufficiently to lift that degeneracy, the rapid expansion of the accreted envelope leads to the ejection of a portion of the envelope. The system then enters a period of quasi-stable nuclear burning on the WD surface until the available fuel source is exhausted \cite{1978A&A....62..339P}. 

The gross observable features of novae are powered by the nuclear burning. Until recently (see \cite{2017NatAs...1..697L}), it was understood that the ejected and expanding envelope acts solely as a photon convertor. The rapid rise in optical luminosity is driven by the expansion of an optically thick envelope. The slower optical decline from peak is linked to the decreasing optical depth of the then optically thin envelope.  Subtleties of the observed features relate to factors such as shocks, ejecta photoionisation and recombination, geometry, inclination, and molecule/dust formation. See \cite{2012BASI...40..185S} for an overview.

Novae are inherently recurrent in nature. A nova eruption does not irreversibly impact the system as a whole --- although the accretion disk is understood to be disrupted or destroyed in most cases (see, for e.g., \cite{2010ApJ...720L.195D} versus \cite{2018ApJ...857...68H}). Post-eruption, the disk will recover or reform \cite{2011ApJ...742..113S}, accretion will resume \cite{2007MNRAS.379.1557W}, and a second eruption will occur once the WD has again accreted a critical mass of material. Theoretical work suggests that recurrence periods range from $50\,\mathrm{d}<P_\mathrm{rec}\lesssim10^{6}$\,yrs \cite{2014ApJ...793..136K,2016ApJ...819..168H,2005ApJ...623..398Y}. A value of a century arbitrarily delimitates the nova population into classical novae (CNe; $P_\mathrm{rec}>100$\,yrs) and recurrent novae (RNe; $P_\mathrm{rec}<100$\,yrs).

This article provides a summary of an invited review talk on the topic of `Recurrent Novae', presented at The Golden Age of Cataclysmic Variables and Related Objects V meeting in Palermo, Italy (Sep.\ 2019)\footnote{A recording of the review talk, and the accompanying slides, may be downloaded from \url{http://www.ljmu.ac.uk/~mjd/talks.html}}. In Section~\ref{rne} RNe are explored in the context of the `MMRD' relationship. In Section~\ref{he-flash}, the reinvigorated link between novae and type Ia SNe is discussed. In Sections~\ref{grne}, \ref{exgal}, and \ref{rrne} the Galactic, extragalactic, and the new population of `rapid recurrent novae' are presented. Finally, in Section~\ref{12a} the most remarkable and extreme nova yet discovered is featured. It is hoped that this article compliments a pair of recent comprehensive reviews of extragalactic novae \cite{10.1088/2514-3433/ab2c63,DARNLEY2019}.

\section{Recurrent Novae and the MMRD}\label{rne}

Any nova that has been observed in eruption more than once is referred to as recurrent. Observed recurrence periods currently range between $1\le P_\mathrm{rec}\le 98$\,yrs \cite{2014A&A...563L...9D,2009AJ....138.1230P}. Using M\,31 and the Milky Way as examples, the RN population only accounts for a few percent of the known nova population \cite{2015ApJS..216...34S}. However, there are clear and substantial selection effects at play with regard to both the RN population size and the range of observed periods \cite{DARNLEY2019}. For example, the RN population of the extensively observed Large Magellanic Cloud (LMC) makes up a much larger proportion of the host's overall nova population \cite{2013AJ....145..117S}. 

The short recurrence periods of the RNe are driven by a {\it combination} of two fundamental system parameters. (Most) RNe contain high mass WDs and have high mass accretion rates ($\dot{M}$). The increased surface gravity of a high mass WD means less material is required to meet the ignition criteria for H-burning. The higher $\dot{M}$ the more rapidly that critical mass is acquired. 

A high mass WD in a nova system can be acquired by two routes:\ the zero-age WD mass was high, the WD is therefore most likely to be ONe in composition \cite{1985ApJS...58..661I}; or the zero-age WD mass was low(er) and the WD mass has increased over the lifetime of the nova system; here the WD is more likely to be CO (see Section~\ref{he-flash}). The high $\dot{M}$ is achieved via Roche-lobe overflow of a sub-giant or through the wind of a red giant donor (the symbiotic case) \cite{2012ApJ...746...61D}, with $\dot{M}$ reaching $\sim10^{-7}$\,M$_\odot$\,yr$^{-1}$. Such systems therefore have orbital periods greater than those of typical CNe ($P_\mathrm{orb}\gtrsim1$\,d versus $P_\mathrm{orb}\sim$ a few hours for CNe  \cite{2012ApJ...746...61D}). Although, in the most extreme case, mass loss may occur through Roche-lobe overflow of a red giant (see Section~\ref{12a}). The lower ignition masses required by RNe leads to substantially less massive, and hence faster evolving, ejecta than are seen from CNe.

\begin{figure}
\begin{center}
\includegraphics[width=\textwidth]{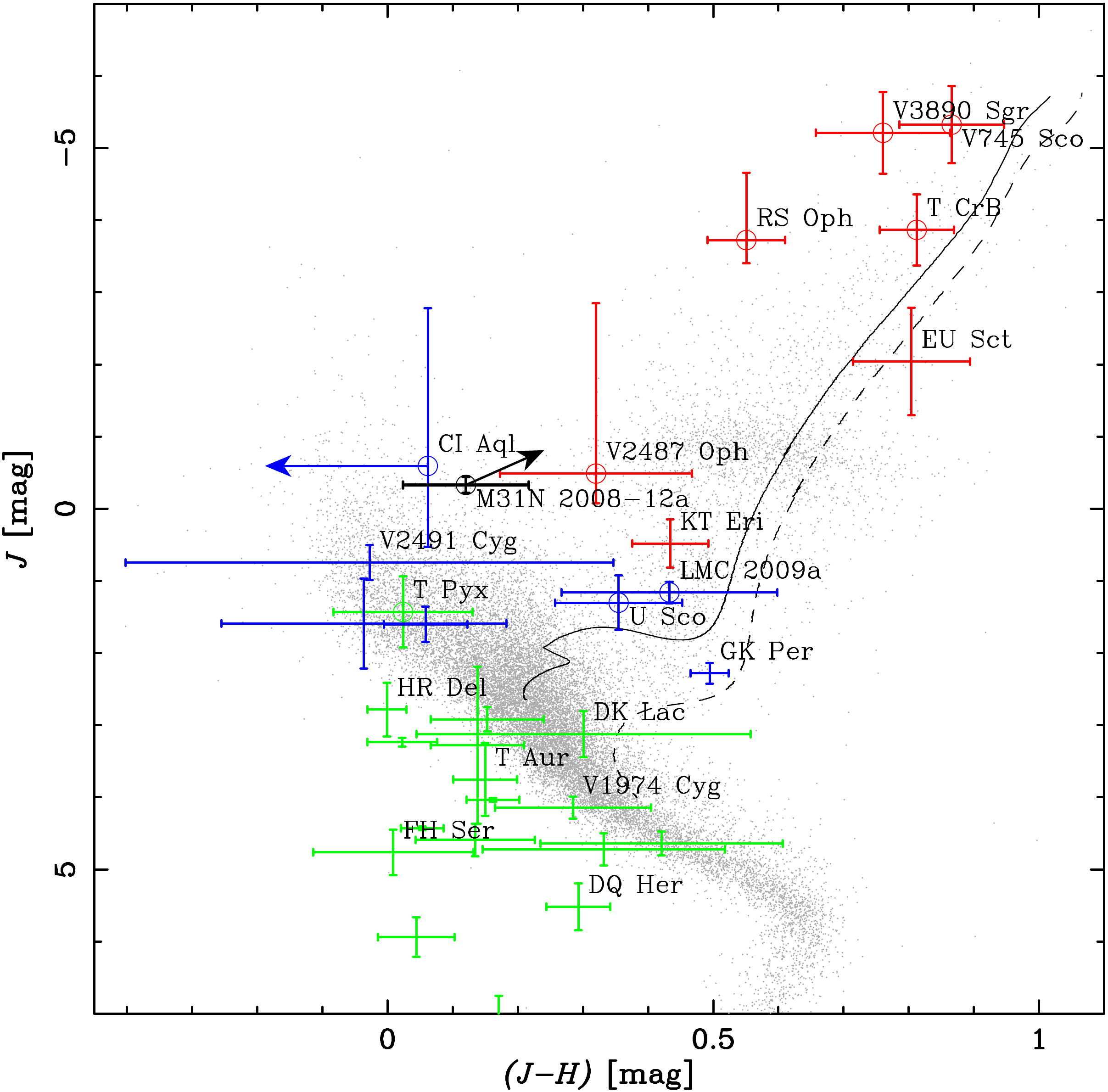}
\end{center}
\caption{Near-infrared colour--magnitude diagram showing the position of selected (mostly) Galactic novae at quiescence. Red data points indicate those with red giant donors, dark-blue those with sub-giant donors, and green those with main sequence donors. Points additionally circled indicate known recurrent novae. The black data point denotes the position of M31N\,2008-12a (see Section~\ref{12a}), with the arrow indicating the evolution during quiescence. The grey data points indicate stars with parallax and photometric errors $<10\%$ extracted from the {\it Hipparcos} catalogue \cite{1997A&A...323L..49P} cross-correlated with 2MASS \cite{2006AJ....131.1163S}, and are included for informative purposes only. The dashed black line and the solid black lines represent evolutionary tracks of 1 and 1.4\,M$_\odot$ Solar composition stars, respectively \cite{2004ApJ...612..168P}. An updated version of Figure~1 from \cite{2012ApJ...746...61D}.\label{colmag}}
\end{figure}

The majority of Galactic RNe follow the picture painted above, with 30\% hosting a sub-giant donor and 50\% with a red giant donor \cite{2012ApJ...746...61D}. Yet two systems, T\,Pyxidis and IM\,Normae --- the ``recurrent unusual novae'' --- seemingly with main sequence donors, lower mass WDs, and short orbital periods, may buck this trend (for a non-complete selection of discussion articles about these unusual RNe see \cite{2010ApJ...708..381S,2017MNRAS.466..581P,2018ApJ...862...89G}). Figure~\ref{colmag} presents a colour-magnitude diagram (based on \cite{2012ApJ...746...61D}) that illustrates how near-IR quiescent photometry is used to determine the donor type in a nova system.

The Maximum Magnitude---Rate of Decline (MMRD) relationship for novae has been with us for almost a century \cite{1929ApJ....69..103H,1945PASP...57...69M}. Simply, the MMRD states that the brighter the nova (at peak) the more rapidly it fades (optically), and typically takes the form of a power-law between maximum luminosity and decay-rate. For a time, the MMRD enabled the popularisation of novae as extragalactic distance indicators --- a potential rival to Cepheid variables. But the large MMRD scatter and the high observational overhead hindered its uptake. More recently, the very concept of the MMRD relationship has been questioned. But even the much heralded {\it Gaia} (DR2; \cite{2018A&A...616A...1G}) has yet been unable to provide compelling evidence either way for Galactic novae (see \cite{2018MNRAS.481.3033S} versus \cite{2019A&A...622A.186S}).

Further afield evidence slowly mounts against the MMRD relationship. First in M\,31 \cite{2011ApJ...735...94K}, and then M\,87 \cite{2017ApJ...839..109S}, a new population of `faint--fast' novae has been proposed --- novae that populate a previously devoid area of the MMRD phase space. Well established nova eruption models (e.g., \cite{2005ApJ...623..398Y}) allow us to interpret the positioning of a nova in the MMRD phase space. The traditional MMRD relationship spans from bright--fast novae that contain high mass WDs accreting at a low rate, to faint--slow novae with low mass WDs and high $\dot{M}$ \cite{DARNLEY2019}\footnote{We note that reliable direct measurements or estimates of the WD mass are only available for a small proportion of all novae. In general the WD mass is inferred from observational properties such as the speed class and independently from the X-ray properties.}. The faint--fast novae, therefore, arise from high mass WDs with high $\dot{M}$. Many of the eight long orbital period Galactic RNe (see Section~\ref{grne}) can be classified as faint--fast novae, but to-date, only one (see Section~\ref{12a}) of their faint--fast M\,31 and M\,87 counterparts has been observed to recur. Recently, \cite{DARNLEY2019} proposed that the apparently empty bright--slow MMRD phase space --- expected to be occupied by low mass WDs with low $\dot{M}$ --- might be occupied by `SPRITES' (eSPecially Red Intermediate-luminosity Transient Events; \cite{2017ApJ...839...88K}), which populate the infrared luminosity space between novae and SNe. This led \cite{DARNLEY2019} to propose that exploration of a `Bolometric MMRD' could have the potential to restore the usefulness of the MMRD.

\section{Helium Flashes and SN\,Ia Progenitors}\label{he-flash}

Nova have long been proposed as a potential progenitor pathway along the single-degenerate route toward SN\,Ia explosions (see, for e.g., \cite{1973ApJ...186.1007W,1999ApJ...519..314H,1999ApJ...522..487H,2000ARA&A..38..191H}). But two barriers have traditionally stood in their way. 

The first is simply whether there are enough novae to make a significant enough contribution to the menagerie of potential SN\,Ia progenitors --- and this remains an open question (see, for e.g., \cite{2015ApJS..216...34S,2014ApJ...788..164P,2014ApJS..213...10W,2016ApJ...817..143W}). Yet novae are probably the most luminous of all pre-SN\,Ia systems, possibly even during quiescence. Thus nova populations can be studied well beyond the heavily biased constraints of the Milky Way, and the size of their contribution to the SN\,Ia rate, and any dependance upon host morphology or star formation history, can in principle be determined observationally.

The second barrier relates to the WD mass and how it changes with a nova eruption. For an accreting WD to reach the Chandrasekhar mass it must, of course, increase in mass with time. Each nova eruption can only ever eject a proportion ($<\!100\%$) of the accreted envelope. Some of the accreted hydrogen must remain on the surface to fuel the on-going nuclear burning that is observed as the super-soft X-ray source (SSS). However, if a large quantity of the WD core material mixes with the accreted material (see, for e.g., \cite{2018A&A...619A.121C}) the ejecta mass can, in principle, be larger than the critical (ignition) mass --- causing the WD mass to decrease. The most recent theoretical studies of nova eruptions --- or H-flashes --- agree that the accretion efficiency $\eta$ over the course of an eruption cycle is $>\!0$ (see, for e.g., \cite{2019arXiv191000575S,2016ApJ...819..168H,2005ApJ...623..398Y,1994ApJ...424..319K,1995ApJ...445..789P,2012BASI...40..419S,2015MNRAS.446.1924H,2015ApJ...808...52K}), i.e.\ the WD mass increases from one H-flash to the next.

Each H-flash deposits its {\it waste products} --- in this case He --- onto the WD core (see, for e.g., \cite{2017ApJ...844..143K}). As time progresses, this leads to a growing He layer on the WD. Just as for the accreted H layer, once the He layer attains enough mass, nuclear burning (via the triple-$\alpha$ process) will ensue within the He layer. Initial simulations of such He-flashes indicated that they violently ejected all accreted material from the WD usually along with a substantial proportion of mixed in core material (see, e.g., \cite{2013MNRAS.433.2884I,2014ASPC..490..287N}. Whether H-flashes allowed WD growth or not, it seemed the He-flashes ultimately whittled away the WD. 

But, the first study to simulate a long series of He-flashes (each punctuated by $\sim$100 H-flashes) made an important discovery \cite{2016ApJ...819..168H}. When the WD core (and He layer) temperature was tracked through the evolution of these flashes it was seen to increase \cite{2016ApJ...819..168H,HILLMAN2019}. After a number of He-flashes the core temperature had increased sufficiently to, at first, lessen the degeneracy of the He layer, and ultimately remove it. As such, these later He-flashes did not result in TNRs and would not eject material from the system. Instead, the later He-flashes simply convert He to C/O and deposit that directly on the WD core \cite{2016ApJ...819..168H,HILLMAN2019}. Ultimately, the WD mass does grow with time --- although initially it may decrease until the WD is sufficiently heated. The potential of novae, and particularly the RNe, as SN\,Ia progenitors was re-ignited.

On this subject, we also refer the reader to the contribution in these proceedings from Sumner Starrfield and collaborators. 

\section{Galactic Recurrent Novae}\label{grne}

The ten {\it known} Galactic RNe are by far (with the exception of perhaps M31N\,2008-12a; see Section~\ref{12a}) the best studied. An extremely comprehensive compilation of their observational properties and history was published in \cite{2010ApJS..187..275S}. The Galactic RNe naturally separate into three distinct classes: the symbiotic RNe, the five (four confirmed; V2487\,Ophiuchi suspected) with red giant donors, often referred to as the {\it RS\,Ophiuchi-class}; the three with sub-giant donors, or the {\it U\,Scorpii-class}; and the {\it T\,Pyxidis}-class, those with short CN-like orbital periods. We summarise some of the properties of these novae in Table~\ref{grn}, which draws from data published in \cite{2012ApJ...746...61D,DARNLEY2019,2014ApJ...788..164P,2010ApJS..187..275S,2008ASPC..401...31A}.

\afterpage{
\clearpage
\thispagestyle{empty}
\begin{landscape}
\centering 
\captionof{table}{Selected properties of the ten known Galactic recurrent novae.\label{grn}}
\begin{threeparttable}
\begin{tabular}{lp{0.37\linewidth}lll}
\hline
Name & Known eruptions$^\mathrm{a}$ & $P_\mathrm{rec~}$ [years]$^\mathrm{b}$ & $P_\mathrm{orb}$ [days] & Next eruption$^\mathrm{c}$ \\
\hline
\multicolumn{3}{c}{{\it The RS\,Oph-class}}\\
\hline
T\,Coronae Borealis & 1866, 1946 & $\sim80$ & 227.57 & $\sim2022$\\
RS\,Ophiuchi & 1898, 1907$^\star$, 1933, 1945$^\star$, 1958, 1967, 1985, 2006 & $15\pm6$ & 455.72 & up to 2027\\
V2487\,Ophiuchi$^\mathrm{d}$ & 1900, 1998 & $\sim98$ & \ldots & $\sim2096$\\
V3890\,Sagittarii & 1962, 1990, 2019 & $29\pm1$ & $519.7\pm0.3$ & $2048\pm1$\\
V745\,Scorpii & 1937, ($\sim$1963), 1989, 2014 & $26\pm1$ & $510\pm20$ & $2040\pm1$ \\
\hline
\multicolumn{3}{c}{{\it The U\,Sco-class}}\\
\hline
CI\,Aquillae & 1917, 1941, ($\sim$1968), 2000 & $27\pm4$ & 0.62 & $2027\pm4$\\
V394\,Coronae Australis & 1949, 1987 & $\sim38$ & 1.52 & $\sim2025$ \\
U\,Scorpii & 1863, ($\sim$1873, $\sim$1884, $\sim$1894), 1906, 1917, ($\sim$1927), 1936, 1945, ($\sim$1955), 1969, 1979, 1987, 1999, 2010 & $10\pm1$ & 1.23 & up to 2021 \\
\hline
\multicolumn{3}{c}{{\it The short orbital period systems}}\\
\hline
IM\,Normae & 1920, 2002 & $\sim82$ & 0.10 & $\sim2084$\\
T\,Pyxidis & 1890, 1902, 1920, 1944, 1967, 2011 & $24\pm12$ & 0.08 & $2035\pm12$\\
\hline
\end{tabular}
\begin{tablenotes}
\small
\item This table is based upon data compiled within \cite{2012ApJ...746...61D,DARNLEY2019,2014ApJ...788..164P,2010ApJS..187..275S,2008ASPC..401...31A} and references therein.
\item[a] Suspected eruptions indicated by $\star$; assumed eruptions (by this author) indicated by parentheses. A 1945 eruption of RS\,Oph was long suspected, \cite{2011MNRAS.414.2195A} presented evidence in support that eruption occurring during a seasonal gap.
\item[b] With the exception of T\,Pyx, the {\it average} inter-eruption times are presented based on the assumption that eruptions (i.e.\ $\dot{M}$) are relatively uniformly spaced (see Section~\ref{12a}). For T\,Pyx we simply report the mean {\it observed} $P_\mathrm{rec}$.
\item[c] Predicted dates of next eruptions are based on the assumed $P_\mathrm{rec}$ for each system. At the time of writing, RS\,Oph and U\,Sco are already within their predicted windows. 
\item[d] There are no orbital period data for V2487\,Oph available. Based on quiescent NIR photometry, \cite{2012ApJ...746...61D} identified the donor as a low-luminosity red giant (possibly red-clump) star and hence placed V2487\,Oph in the RS\,Oph class.
\end{tablenotes}
\end{threeparttable}
\end{landscape}
\clearpage
}

The population of known Galactic RNe must clearly suffer from (in many cases) severe selection effects and biases. At the time of writing, the shortest inter-eruption period yet seen is the eight years that elapsed between the 1979 and 1987 eruptions of U\,Sco --- whereas numerous examples of much shorter recurrence periods have been seen extragalactically (see Sections~\ref{exgal} and \ref{rrne}). At the other end of the scale, the longest inter-eruption period is the 98 years between the two known eruptions of V2487\,Oph \cite{2009AJ....138.1230P} --- an upper limit that must surely only be related to the availability of historical observations. In a century's time, I firmly predict that the longest known inter-eruption period will be $\sim$200\,yrs.

While there are only ten confirmed Galactic RNe, there are almost as many strong candidates (see \cite{2014ApJ...788..164P} for a compilation). In all cases these are CNe (a single eruption) that share many observational properties with an eruption of a RN. There are very strong contenders, such as KT\,Eridani \cite{2010ApJ...724..480H,2012A&A...537A..34J,2013MNRAS.433.1991R,2014A&A...564A..76M}; and more `conflicted' cases, such as V2491\,Cygni \cite{2014ApJ...788..164P,2010MNRAS.401..121P,2011MNRAS.412.1701R,2011NewA...16..209M,2011A&A...530A..70D}. All of these systems deserve continued study and warrant long-term monitoring to catch the expected, or proposed, second eruptions. 

In an attempt to quantify the number of RNe that {\it masquerade} as CNe, \cite{2014ApJ...788..164P} compiled a set of observational properties of RNe. These include:\ low amplitude eruptions (due to the high quiescent luminosity of an evolved donor and a high $\dot{M}$ --- therefore bright --- accretion disk); long orbital periods (required to physically accommodate the enlarged radius of an evolved donor); NIR bright at quiescent (again a property of the donor); high velocity ejecta (due to the high surface gravity, and hence high escape velocity, of the WD); high excitation lines (visible due to the energetics of the eruption and the low optical depth of the low mass ejecta); and a light curve plateau (proposed by many, including \cite{2018ApJ...857...68H,2008ASPC..401..206H}, to be due to unveiling of the surviving high $\dot{M}$ accretion disk as the photosphere recedes back toward the WD). A key prediction from \cite{2014ApJ...788..164P} is that around a third of all Galactic novae may be recurrent (formally $10\leq P_\mathrm{rec}\leq100$\,yrs; A.\,Pagnotta, priv.\ comm.).

\section{Extragalactic Recurrent Novae}\label{exgal}

There are currently over 1100 nova candidates within M\,31\footnote{For the most complete, and up to date, catalogue, see:\ \url{http://www.mpe.mpg.de/~m31novae/index.php}} \cite{2010AN....331..187P}. Indeed there are more nova candidates in M\,31 than known novae in all other galaxies (the Milky Way included) combined. By virtue of its high nova eruption rate \cite{2006MNRAS.369..257D} combined with its closeness, M\,31 has become the preferred laboratory for the study of nova populations \cite{DARNLEY2019}. But M\,31 is far from ideal, its large angular size has led to the majority of surveys (particularly historic campaigns) focussing only on the bulge or the brighter inner regions \cite{DARNLEY2019}. This is far from surprising, with the nova population of a galaxy broadly following the light, additional fields away from the centre only provide diminishing returns. Of course this places substantial spatial (and temporal) biases on any population studies using those novae. Viewed from the Milky Way, M\,31 is relative close to being edge-on \cite{1985ApJ...295..287D}, which results in large internal extinction uncertainties \cite{2006MNRAS.369..257D,2015ApJ...814....3D} and severely challenges efforts to disentangle the bulge and disk populations within that host \cite{2006MNRAS.369..257D}. However, \cite{2006MNRAS.369..257D} found significant evidence for separate bulge and disk nova populations within M\,31. More recently \cite{tramp}, utilising wide-field high-cadence surveys of M\,31, reported the discovery of two novae associated with the Giant Stellar Stream to the south of M\,31 (see \cite{2001Natur.412...49I}).

Of those 1100 M\,31 nova candidates, in excess of 200 have now been spectroscopically confirmed \cite{2011ApJ...734...12S,Ransome2019}. Within that subset lie eighteen known RNe. While a small number of the M\,31 RNe have been known or suspected for some time, the first comprehensive catalogue of these systems was published by \cite{2015ApJS..216...34S}, who found 16. An additional three M\,31 RNe have been added \cite{2015ATel.7116....1H,Sch2017,2017ATel11088....1W,2017ATel10001....1S}, and one recanted \cite{2017RNAAS...1a..44S}. A completeness analysis (of the 16) \cite{2015ApJS..216...34S} indicated that the detection efficiency of M\,31 RNe --- i.e. the probability of detecting at least a second eruption --- may be as low as 10\%. Subsequently, as many as 1 in 3 M\,31 nova eruptions may be from a RN (here, formally, $1\leq P_\mathrm{rec}\leq100$\,yrs).

Add in to the mix the four known RNe in the LMC \cite{DARNLEY2019,2013AJ....145..117S} and that's it --- 32 known RNe spanning just three galaxies. In most other Local Group galaxies the observations are just too sparse (e.g., M\,33; \cite{2012ApJ...752..156S}) or the nova rate too low (e.g., SMC; \cite{2013AJ....145..117S,2016ApJS..222....9M}) to have yet reached a good probability of detecting a RN. Beyond the Local Group, M\,81 \cite{2004AJ....127..816N} and M\,87 \cite{2016ApJS..227....1S} are the next best sampled hosts, but they too are yet to produce a published RN.

The first extragalactic nova progenitor, or quiescent system, that of M31N\,2007-12b, was recovered in M\,31 using archival {\it Hubble Space Telescope (HST)} ACS/WFC data \cite{2009ApJ...705.1056B}. That {\it candidate} (still unconfirmed) RN was shown to harbour a red giant donor. Subsequently, \cite{2014ASPC..490...49D} demonstrated that all novae with red giant donors (which Galactically are dominated by known RNe; \cite{2012ApJ...746...61D}; see Figure~\ref{colmag}) are accessible to {\it HST} within the Local Group. The first extragalactic survey for quiescent novae was conducted within M\,31 by \cite{2014ApJS..213...10W} who produced a catalogue of eleven progenitor systems, each hosting a red giant donor. The subsequent statistical analysis \cite{2016ApJ...817..143W} reported that 30\% of M\,31 novae may contain a red giant donor. Moreover, that population of {\it potential} RNe was strongly associated with the (young) disk population of M\,31. The \cite{2016ApJ...817..143W} result was indeed statistically consistent with {\it all} the red giant novae being disk novae. While far from proven, a potential consequence might be that novae (with red giant donors) only contribute to the SN\,Ia rate in late-type galaxies.

\section{Rapid Recurrent Novae}\label{rrne}

As a WD in a nova system approaches the Chandrasekhar mass, $P_\mathrm{rec}$ should decrease (see, for e.g., \cite{2005ApJ...623..398Y}). Theory shows us that on the cusp of breaching that limit $P_\mathrm{rec}$ may be as low as two months \cite{2014ApJ...793..136K,2016ApJ...819..168H}. But when we look at the Galactic RN population the most rapidly recurring system is U\,Sco with a mean inter-eruption time scale of {\it only} 10\,yrs --- close to two orders of magnitude slower than the predicted extreme. So if a nova WD really can be grown toward the Chandrasekhar mass, where are the more rapidly recurring systems?

\begin{figure}
\begin{center}
\includegraphics[width=\textwidth]{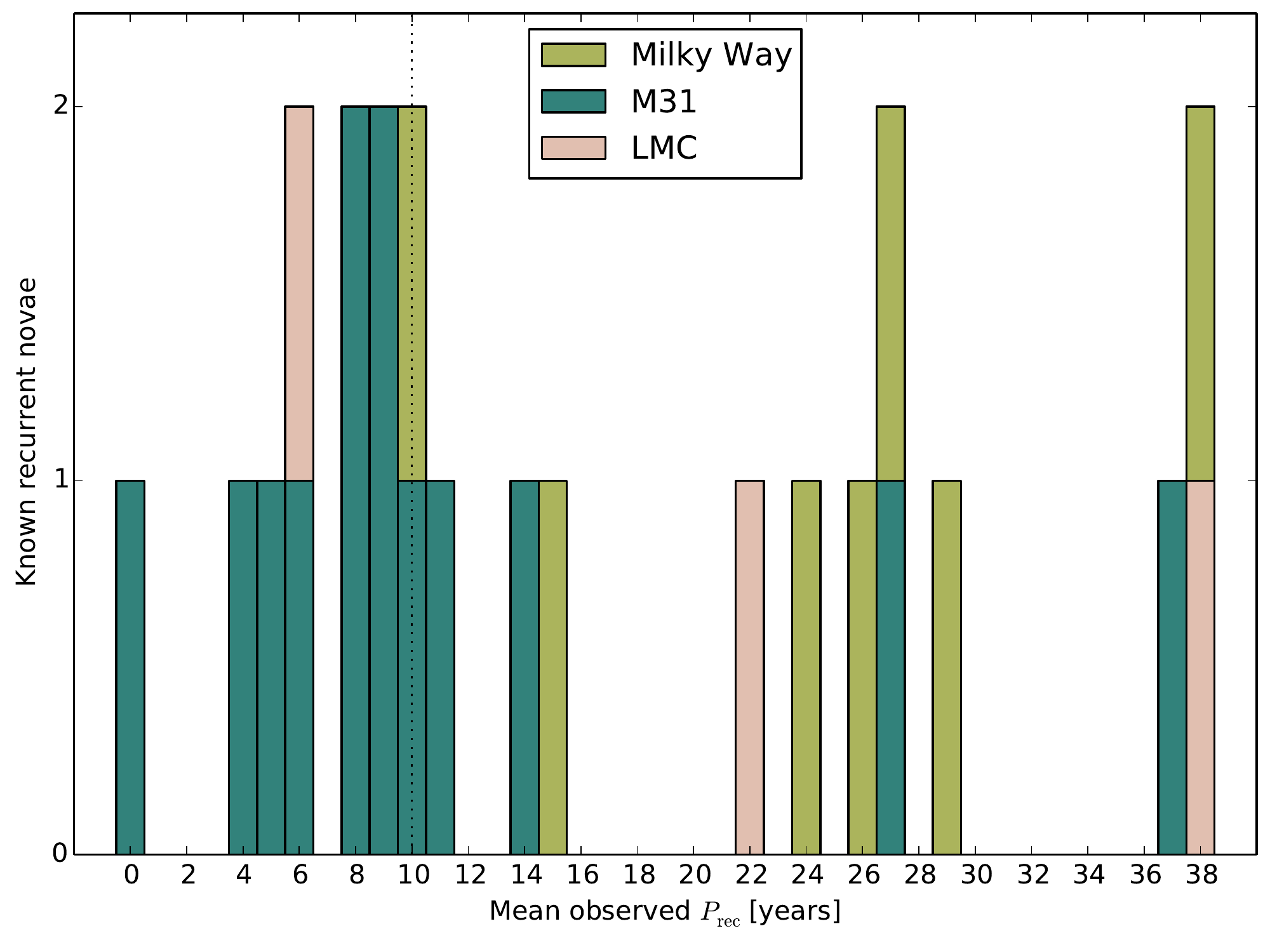}
\end{center}
\caption{The distribution of the mean inter-eruption periods of the known recurrent novae (the plot only shows systems with $P_\mathrm{rec}\leq40$\,yrs; 9 RNe with longer periods are not shown). The dotted vertical line demarks the realm of the rapid recurrent novae ($P_\mathrm{rec}\leq10$\,yrs). Adapted and updated from Figure~3 of \cite{DARNLEY2019}.\label{rrne_p}}
\end{figure}

If one examines the RN populations of the LMC and M\,31 there is a striking difference between the RNe from those hosts and those from the Milky Way. Of the four known LMC RNe, one (LMCN\,1968-12a) has $P_\mathrm{rec}=6.2\pm1.2$\,yrs \cite{2019MNRAS.tmp.2562K} --- almost twice as {\it fast} as U\,Sco. But of the eighteen M\,31 RNe (see Section~\ref{exgal}) {\it half} have recurrence periods similar to or shorter than U\,Sco! The existence of this {\it quick-fire} population of RNe led \cite{DARNLEY2019} to dub all systems with $P_\mathrm{rec}\leq10$\,yrs `rapid recurrent novae' (RRNe). In fact all new M\,31 RNe discovered since 1984 --- of which there are seven --- are RRNe. Figure~\ref{rrne_p} illustrates the distribution of RN recurrence periods within the Milky Way, M\,31, and the LMC. Is this difference due to variation between the stellar (and therefore nova) populations of these galaxies, related possibly to differing star formation histories and metallicities, or are  observational select effects at play?

So where are the Galactic RRNe? {\it If they exist}, they are most likely hiding in plain sight. RRNe must contain high mass WDs accreting at an elevated rate \cite{2005ApJ...623..398Y}. Thus RRNe should be faint--fast novae (see Section~\ref{rne}). RRNe are therefore particularly challenging to detect, let alone classify. Galactically, perhaps somewhat ironically, the largest background signal to RRNe may actually be the dwarf novae (DNe; see \cite{1995cvs..book.....W} for an authoritative overview). Unfortunately, most DNe are not followed-up sufficiently (i.e.\ optical spectroscopy and X-ray observations, and in many cases not even photometrically) to be able to distinguish them from a low amplitude, quasi-short-periodic, nova eruption --- a RRN eruption.

But we have been able to successfully detect and follow-up RRNe beyond the Milky Way, particularly in M\,31 (e.g., \cite{2014A&A...563L...9D}), because M\,31 (and the LMC) cover relatively small regions of the sky. The advent of large area detectors coupled with high cadence observations, and the invaluable input from the amateur astronomy community, have enabled M\,31 to be covered in both unprecedented depth and cadence over the last decade, or so. This change in observing strategy and capability led directly to the discovery of RRNe. {\it Rapid response} follow-up facilities, such as the Liverpool Telescope \cite{2004SPIE.5489..679S} and the {\it Neil Gehrels Swift Observatory} \cite{2004ApJ...611.1005G,PAGE2019}, have permitted the study and confirmation of those RRN candidates\footnote{It may also be worthwhile noting that (M\,31) DNe are not visible above from unresolved surface brightness of M\,31 with the typical seeing experienced by ground-based observations.}.

We stand on the precipice of a new era of all-sky high cadence surveys. The SSS-phase of RRNe (and novae in general) is perhaps their most {\it unique} `fingerprint'. The recently launched {\it eROSITA} (see, for e.g., \cite{2012arXiv1209.3114M}) could be vital for detecting new RRNe in the Milky Way. Optical facilities, such as the Large Synoptic Survey Telescope (LSST; \cite{2019ApJ...873..111I}) should dramatically alter the landscape for transient object astronomy. LSST, for example, will be sensitive to RRN eruptions in the Milky Way and Magellanic Clouds. But for optical discoveries of RRNe, the problem will ultimately lie with the availability of fast enough spectroscopic (or soft X-ray) follow-up. While many follow-up facilities complementing LSST and its contemporaries have been or are being planned and built (e.g., \cite{2015ExA....39..119C}) there are unlikely to be anywhere near enough to classify all new transient detections. So for any novae, not specifically RRNe, to reap any benefit from LSST et al., the value of nova science to all of astronomy requires substantial promotion. Right now, one RRN system in particular may be key.

\section{M31N\,2008-12a}\label{12a}

``In all your travels, have you ever seen a star go supernova?'' \cite{BSG}. Any denizen of the Andromeda Galaxy, particularly the region surrounding the unassumingly named binary system RX\,J0045.4+4154, may have borne witness to a nearby star undergoing a fantastic increase in luminosity around once a (Terran) year. Perhaps this system, may even be central to the culture of those nearby `Andromedans'. Little might they know that, given the right conditions, and the right composition, RX\,J0045.4+4154 may one day explode as a SN\,Ia, and potentially --- unfortunately --- wipe out all in the immediate vicinity. 

RX\,J0045.4+4154 itself usually goes by its more familiar (yet equally cumbersome) name M31N\,2008-12a --- it is the prototype RRN, and the most extreme example of a nova yet discovered. Of all known novae, probably of all known stellar systems, is the most likely to explode {\it first} as a SN\,Ia. And, given the distance to M\,31, it probably already has.

The first observed eruption of `12a' (by humans) took place on Christmas Day 2008 \cite{2008Nis}. Since then, it has been observed in eruption {\it every single year}, most recently in November 2019 \cite{2019ATel13269....1O}. Although suspected beforehand, the true nature of 12a was not confirmed until the 2013 eruption \cite{2014A&A...563L...9D,2014A&A...563L...8H,2014ApJ...786...61T}. At the time, annual nova eruptions were completely unprecedented. Although a small number of RRNe had already been found in M\,31, none had been studied in any detail. Until then the most rapidly recurring (and therefore extreme) RN was generally accepted to be U\,Sco.

12a is {\it powered} by the combination of the most massive WD known (predicted via some models to be $\simeq1.38\,\mathrm{M}_\odot$; \cite{2015ApJ...808...52K}) accreting, stably, at an exceptionally high rate in the region of $\left(0.6\lesssim \dot{M} \lesssim1.4\right)\times10^{-6}\,\mathrm{M}_\odot\,\mathrm{yr}^{-1}$ \cite{2017ApJ...849...96D}. Together, this drives eruptions with a {\it mean} $P_\mathrm{rec}=0.99\pm0.02$\,yrs \cite{DARNLEY2019}. The high accretion rate has been proposed to be due to a Roche-lobe overflowing red giant (or red clump) donor \cite{2017ApJ...849...96D}. At present, the composition of the 12a WD has not been confirmed. But, {\it HST} STIS FUV spectra taken just 3\,d after the 2015 eruption showed no indication of Ne in the ejecta\footnote{A Ne {\it over-abundance} is a signature of an ONe WD \cite{1985MNRAS.212..753W,1986ApJ...303L...5S}.} \cite{2017ApJ...847...35D}. This {\it implies} --- {\it but does not confirm} --- that the 12a WD is CO and that the system is building toward a SN\,Ia explosion. A very conservative prediction suggested such an event will occur in $<20$\,kyr \cite{2017ApJ...849...96D}.

Comprehensive overviews and summaries of 12a and its eruptions are presented in \cite{DARNLEY2019,2016ApJ...833..149D,2017ASPC..509..515D}. Selected highlights, that illustrate the extreme nature of the system include:\ low mass ejecta with low {\it optical} luminosity, consistent with high {$\dot{M}$, high mass WD, and ultra-short $P_\mathrm{rec}$ models of TNRs \cite{2005ApJ...623..398Y,2015ApJ...808...52K}; ejecta deceleration as they shock the pre-existing disk/donor wind \cite{2017ApJ...849...96D,2016ApJ...833..149D}; the highest SSS effective temperature seen in a nova ($T_\mathrm{eff}\sim120$\,eV; \cite{2016ApJ...833..149D}); detection of significant high-amplitude SSS variability \cite{2018ApJ...857...68H}; remarkable similarity between the 2013--2015 and 2017--2019 eruptions \cite{2016ApJ...833..149D,12a1718}; detection of short-lived high velocity, collimated {\it(jet-like)}, outflows \cite{2017ApJ...847...35D,2016ApJ...833..149D}; and recovery of the progenitor from archival {\it HST} data \cite{2014A&A...563L...9D,2017ApJ...849...96D}.

But the most remarkable aspect of 12a is the effect it has had upon the local environment. The system is surrounded by a vast shell-like nebula that is the remnant of all the eruptions that have enabled the WD to grow toward the Chandrasekhar mass \cite{2015A&A...580A..45D,2019Natur.565..460D}. This `nova super-remnant' (NSR) has a major axis of 134\,pc and consists of $\sim10^{5-6}\,\mathrm{M}_\odot$ of swept up ISM \cite{2019Natur.565..460D}. Hydrodynamical simulations of 100,000 periodic (annual) 12a-like eruptions clearly demonstrated that repeated RN eruptions are capable of producing such a phenomenon \cite{2019Natur.565..460D}.

Any nova system, as it grows toward the Chandrasekhar mass, should therefore be surrounded by a similar NSR. It is clear that the size of these objects will depend upon parameters such as the zero-age and current WD mass, the accretion history, and surrounding ISM density. But factors such as metalicity, donor type, and even line-of-sight inclination may affect the observational properties. NSRs clearly deserve substantial further theoretical and observational study. The hunt is already on to find further examples.

\section{Summary}

As a sub-set of all novae, RNe present generally the more extreme system and eruption parameters. Their short inter-eruption periods permit detailed study of repeated eruptions from the same system. With growing WDs, RNe are strong candidates for a single-degenerate SN\,Ia pathway. In general, the shorter the recurrence period the closer a given system is to reaching the Chandrasekhar mass. Currently, the known RN population is beset by substantial selection effects, particularly Galactically. But those challenges can be, and are being, overcome. By doing so the contribution of novae to the SN\,Ia rate can be quantified.

Recurrent nova super-remnants present a new opportunity to not only identify previously unknown RNe and RRNe, but potentially to locate any `extinct' systems --- former RNe, where the donor mass reservoir has been depleted. But perhaps more intriguingly, should a RN explode as a SN\,Ia --- due to its vast scale  --- the NSR will provide a clear {\it and persistent} signpost to the progenitor-type of that SN\,Ia. By its very nature, NSR formation will have also pre-prepared the local environment in which the SN\,Ia will explode. Importantly, if the 12a NSR is common, such RN eruptions will have moved around $10^6\,\mathrm{M}_\odot$ of (predominantly) hydrogen away from the site of the imminent SN\,Ia, and it will take centuries before an expanding SN\,Ia shock interacts with that material. RNe and NSRs may therefore provide a solution to the single-degenerate SN\,Ia `paradox'; where is the hydrogen?

Of course, the big unanswered question is whether M31N\,2008-12a and its super-remnant is a rare, or even unique, phenomenon; or is it just the tip of the iceberg?

\acknowledgments

The study of novae, Galactically and extragalactically, is heavily indebted to the continuing support provided by the amateur astronomy community. I would therefore like to take this opportunity to pass on my thanks to all those who contribute to the work of the nova community, and astronomy in general. That includes those affiliated with the AAVSO, the BAA, and the VSOLJ, from which the 12a work in particular draws heavily. I would like to thank the Scientific and Local Organising Committees of The Golden Age of Cataclysmic Variables and Related Objects V, particularly Franco Giovannelli and Francesco Reale, for their kind invitation to present a review talk on Recurrent Novae. I would also like to extend my thanks to the organisers for still being able to accommodate me (remotely) once it became clear that I was unable to travel to the workshop. I would also like to thank my friend and collaborator Martin Henze for useful comments on this manuscript. I must acknowledge partial funding from the UK Science and Technology Facilities Council (STFC), who supported much of my own work over the past six years. Finally, I would like to pass my thanks to the referee of this review, Sumner Starrfield, whose comments help improved this work.

\bibliographystyle{JHEP}
\bibliography{refs.bib}

\providecommand{\href}[2]{#2}\begingroup\raggedright\begin{thebibliography}{100}

\bibitem{2011PhDT........35K}
M.~M. {Kasliwal}, \emph{{Bridging the gap : elusive explosions in the local
  universe}}, Ph.D. thesis, California Institute of Technology, Jan, 2011.

\bibitem{2017ApJ...834..196S}
A.~W. {Shafter}, \emph{{The Galactic Nova Rate Revisited}},
  \href{https://doi.org/10.3847/1538-4357/834/2/196}{\emph{\apj} {\bfseries
  834} (2017) 196} [\href{https://arxiv.org/abs/1606.02358}{{\ttfamily
  1606.02358}}].

\bibitem{2006MNRAS.369..257D}
M.~J. {Darnley}, M.~F. {Bode}, E.~{Kerins}, A.~M. {Newsam}, J.~{An},
  P.~{Baillon} et~al., \emph{{Classical novae from the POINT-AGAPE microlensing
  survey of M31 - II. Rate and statistical characteristics of the nova
  population}},
  \href{https://doi.org/10.1111/j.1365-2966.2006.10297.x}{\emph{\mnras}
  {\bfseries 369} (2006) 257}
  [\href{https://arxiv.org/abs/astro-ph/0509493}{{\ttfamily
  astro-ph/0509493}}].

\bibitem{2016ApJS..227....1S}
M.~M. {Shara}, T.~F. {Doyle}, T.~R. {Lauer}, D.~{Zurek}, J.~D. {Neill}, J.~P.
  {Madrid} et~al., \emph{{A Hubble Space Telescope Survey for Novae in M87. I.
  Light and Color Curves, Spatial Distributions, and the Nova Rate}},
  \href{https://doi.org/10.3847/0067-0049/227/1/1}{\emph{\apjs} {\bfseries 227}
  (2016) 1} [\href{https://arxiv.org/abs/1602.00758}{{\ttfamily 1602.00758}}].

\bibitem{2017RNAAS...1a..11S}
A.~W. {Shafter}, A.~{Kundu} and M.~{Henze}, \emph{{On the Nova Rate in M87}},
  \href{https://doi.org/10.3847/2515-5172/aa9847}{\emph{Research Notes of the
  American Astronomical Society} {\bfseries 1} (2017) 11}
  [\href{https://arxiv.org/abs/1712.05818}{{\ttfamily 1712.05818}}].

\bibitem{2015Natur.518..381T}
A.~{Tajitsu}, K.~{Sadakane}, H.~{Naito}, A.~{Arai} and W.~{Aoki},
  \emph{{Explosive lithium production in the classical nova V339 Del (Nova
  Delphini 2013)}}, \href{https://doi.org/10.1038/nature14161}{\emph{\nat}
  {\bfseries 518} (2015) 381}
  [\href{https://arxiv.org/abs/1502.05598}{{\ttfamily 1502.05598}}].

\bibitem{2018MNRAS.478.1601I}
L.~{Izzo}, P.~{Molaro}, P.~{Bonifacio}, M.~{Della Valle}, Z.~{Cano}, A.~{de
  Ugarte Postigo} et~al., \emph{{Beryllium detection in the very fast nova
  ASASSN-16kt (V407 Lupi)}},
  \href{https://doi.org/10.1093/mnras/sty435}{\emph{\mnras} {\bfseries 478}
  (2018) 1601} [\href{https://arxiv.org/abs/1802.05896}{{\ttfamily
  1802.05896}}].

\bibitem{2019arXiv191000575S}
S.~{Starrfield}, M.~{Bose}, C.~{Iliadis}, W.~R. {Hix}, C.~E. {Woodward} and
  R.~M. {Wagner}, \emph{{Carbon-Oxygen Classical Novae are Galactic $^7$Li
  Producers as well as Potential Supernova Ia Progenitors}}, {\emph{arXiv
  e-prints} (2019) arXiv:1910.00575}
  [\href{https://arxiv.org/abs/1910.00575}{{\ttfamily 1910.00575}}].

\bibitem{Jos16}
J.~{Jos\'e}, \emph{{Stellar Explosions: Hydrodynamics and Nucleosynthesis}}.
  CRC/Taylor and Francis, Boca Raton, FL, USA, 2016,
  \href{https://doi.org/10.1201/b19165}{10.1201/b19165}.

\bibitem{2012ApJ...746...61D}
M.~J. {Darnley}, V.~A.~R.~M. {Ribeiro}, M.~F. {Bode}, R.~A. {Hounsell} and
  R.~P. {Williams}, \emph{{On the Progenitors of Galactic Novae}},
  \href{https://doi.org/10.1088/0004-637X/746/1/61}{\emph{\apj} {\bfseries 746}
  (2012) 61} [\href{https://arxiv.org/abs/1112.2589}{{\ttfamily 1112.2589}}].

\bibitem{2008clno.book.....B}
M.~F. {Bode} and A.~{Evans}, eds., \emph{{Classical Novae, 2nd Edition}},
  vol.~43 of \emph{Cambridge Astrophysics Series}. Cambridge University Press,
  Cambridge, Apr., 2008.

\bibitem{2014ASPC..490.....W}
P.~A. {Woudt} and V.~A.~R.~M. {Ribeiro}, eds., \emph{{Stella Novae: Past and
  Future Decades}}, vol.~490 of \emph{Astronomical Society of the Pacific
  Conference Series}, (San Francisco), Astronomical Society of the Pacific,
  Dec., 2014.

\bibitem{1972ApJ...176..169S}
S.~{Starrfield}, J.~W. {Truran}, W.~M. {Sparks} and G.~S. {Kutter}, \emph{{CNO
  Abundances and Hydrodynamic Models of the Nova Outburst}},
  \href{https://doi.org/10.1086/151619}{\emph{\apj} {\bfseries 176} (1972)
  169}.

\bibitem{2016PASP..128e1001S}
S.~{Starrfield}, C.~{Iliadis} and W.~R. {Hix}, \emph{{The Thermonuclear Runaway
  and the Classical Nova Outburst}},
  \href{https://doi.org/10.1088/1538-3873/128/963/051001}{\emph{\pasp}
  {\bfseries 128} (2016) 051001}.

\bibitem{1978A&A....62..339P}
D.~{Prialnik}, M.~M. {Shara} and G.~{Shaviv}, \emph{{The evolution of a slow
  nova model with a Z = 0.03 envelope from pre-explosion to extinction}},
  {\emph{\aap} {\bfseries 62} (1978) 339}.

\bibitem{2017NatAs...1..697L}
K.-L. {Li}, B.~D. {Metzger}, L.~{Chomiuk}, I.~{Vurm}, J.~{Strader},
  T.~{Finzell} et~al., \emph{{A nova outburst powered by shocks}},
  \href{https://doi.org/10.1038/s41550-017-0222-1}{\emph{Nature Astronomy}
  {\bfseries 1} (2017) 697} [\href{https://arxiv.org/abs/1709.00763}{{\ttfamily
  1709.00763}}].

\bibitem{2012BASI...40..185S}
S.~N. {Shore}, \emph{{Spectroscopy of novae -- a user's manual}},
  {\emph{Bulletin of the Astronomical Society of India} {\bfseries 40} (2012)
  185} [\href{https://arxiv.org/abs/1211.3176}{{\ttfamily 1211.3176}}].

\bibitem{2010ApJ...720L.195D}
J.~J. {Drake} and S.~{Orlando}, \emph{{The Early Blast Wave of the 2010
  Explosion of U Scorpii}},
  \href{https://doi.org/10.1088/2041-8205/720/2/L195}{\emph{\apjl} {\bfseries
  720} (2010) L195} [\href{https://arxiv.org/abs/1007.2810}{{\ttfamily
  1007.2810}}].

\bibitem{2018ApJ...857...68H}
M.~{Henze}, M.~J. {Darnley}, S.~C. {Williams}, M.~{Kato}, I.~{Hachisu}, G.~C.
  {Anupama} et~al., \emph{{Breaking the Habit: The Peculiar 2016 Eruption of
  the Unique Recurrent Nova M31N 2008-12a}},
  \href{https://doi.org/10.3847/1538-4357/aab6a6}{\emph{\apj} {\bfseries 857}
  (2018) 68} [\href{https://arxiv.org/abs/1803.00181}{{\ttfamily 1803.00181}}].

\bibitem{2011ApJ...742..113S}
B.~E. {Schaefer}, A.~{Pagnotta}, A.~P. {LaCluyze}, D.~E. {Reichart}, K.~M.
  {Ivarsen}, J.~B. {Haislip} et~al., \emph{{Eclipses during the 2010 Eruption
  of the Recurrent Nova U Scorpii}},
  \href{https://doi.org/10.1088/0004-637X/742/2/113}{\emph{\apj} {\bfseries
  742} (2011) 113} [\href{https://arxiv.org/abs/1108.1214}{{\ttfamily
  1108.1214}}].

\bibitem{2007MNRAS.379.1557W}
H.~L. {Worters}, S.~P.~S. {Eyres}, G.~E. {Bromage} and J.~P. {Osborne},
  \emph{{Resumption of mass accretion in RS Oph}},
  \href{https://doi.org/10.1111/j.1365-2966.2007.12066.x}{\emph{\mnras}
  {\bfseries 379} (2007) 1557}
  [\href{https://arxiv.org/abs/0706.1213}{{\ttfamily 0706.1213}}].

\bibitem{2014ApJ...793..136K}
M.~{Kato}, H.~{Saio}, I.~{Hachisu} and K.~{Nomoto}, \emph{{Shortest Recurrence
  Periods of Novae}},
  \href{https://doi.org/10.1088/0004-637X/793/2/136}{\emph{\apj} {\bfseries
  793} (2014) 136} [\href{https://arxiv.org/abs/1404.0582}{{\ttfamily
  1404.0582}}].

\bibitem{2016ApJ...819..168H}
Y.~{Hillman}, D.~{Prialnik}, A.~{Kovetz} and M.~M. {Shara}, \emph{{Growing
  White Dwarfs to the Chandrasekhar Limit: The Parameter Space of the Single
  Degenerate SNIa Channel}},
  \href{https://doi.org/10.3847/0004-637X/819/2/168}{\emph{\apj} {\bfseries
  819} (2016) 168} [\href{https://arxiv.org/abs/1508.03141}{{\ttfamily
  1508.03141}}].

\bibitem{2005ApJ...623..398Y}
O.~{Yaron}, D.~{Prialnik}, M.~M. {Shara} and A.~{Kovetz}, \emph{{An Extended
  Grid of Nova Models. II. The Parameter Space of Nova Outbursts}},
  \href{https://doi.org/10.1086/428435}{\emph{\apj} {\bfseries 623} (2005) 398}
  [\href{https://arxiv.org/abs/astro-ph/0503143}{{\ttfamily
  astro-ph/0503143}}].

\bibitem{10.1088/2514-3433/ab2c63}
A.~W. Shafter, \emph{Extragalactic Novae}, 2514-3433. IOP Publishing, 2019,
  \href{https://doi.org/10.1088/2514-3433/ab2c63}{10.1088/2514-3433/ab2c63}.

\bibitem{DARNLEY2019}
M.~J. Darnley and M.~Henze, \emph{On a century of extragalactic novae and the
  rise of the rapid recurrent novae},
  \href{https://doi.org/https://doi.org/10.1016/j.asr.2019.09.044}{\emph{Advances
  in Space Research} (2019) }.

\bibitem{2014A&A...563L...9D}
M.~J. {Darnley}, S.~C. {Williams}, M.~F. {Bode}, M.~{Henze}, J.-U. {Ness},
  A.~W. {Shafter} et~al., \emph{{A remarkable recurrent nova in M 31: The
  optical observations}},
  \href{https://doi.org/10.1051/0004-6361/201423411}{\emph{\aap} {\bfseries
  563} (2014) L9} [\href{https://arxiv.org/abs/1401.2905}{{\ttfamily
  1401.2905}}].

\bibitem{2009AJ....138.1230P}
A.~{Pagnotta}, B.~E. {Schaefer}, L.~{Xiao}, A.~C. {Collazzi} and P.~{Kroll},
  \emph{{Discovery of a Second Nova Eruption of V2487 Ophiuchi}},
  \href{https://doi.org/10.1088/0004-6256/138/5/1230}{\emph{\aj} {\bfseries
  138} (2009) 1230} [\href{https://arxiv.org/abs/0908.2143}{{\ttfamily
  0908.2143}}].

\bibitem{2015ApJS..216...34S}
A.~W. {Shafter}, M.~{Henze}, T.~A. {Rector}, F.~{Schweizer}, K.~{Hornoch},
  M.~{Orio} et~al., \emph{{Recurrent Novae in M31}},
  \href{https://doi.org/10.1088/0067-0049/216/2/34}{\emph{\apjs} {\bfseries
  216} (2015) 34} [\href{https://arxiv.org/abs/1412.8510}{{\ttfamily
  1412.8510}}].

\bibitem{2013AJ....145..117S}
A.~W. {Shafter}, \emph{{Photometric and Spectroscopic Properties of Novae in
  the Large Magellanic Cloud}},
  \href{https://doi.org/10.1088/0004-6256/145/5/117}{\emph{\aj} {\bfseries 145}
  (2013) 117} [\href{https://arxiv.org/abs/1302.6285}{{\ttfamily 1302.6285}}].

\bibitem{1985ApJS...58..661I}
J.~{Iben}, I. and A.~V. {Tutukov}, \emph{{On the evolution of close binaries
  with components of initial mass between 3 M and 12 M.}},
  \href{https://doi.org/10.1086/191054}{\emph{\apjs} {\bfseries 58} (1985)
  661}.

\bibitem{2010ApJ...708..381S}
B.~E. {Schaefer}, A.~{Pagnotta} and M.~M. {Shara}, \emph{{The Nova Shell and
  Evolution of the Recurrent Nova T Pyxidis}},
  \href{https://doi.org/10.1088/0004-637X/708/1/381}{\emph{\apj} {\bfseries
  708} (2010) 381} [\href{https://arxiv.org/abs/0906.0933}{{\ttfamily
  0906.0933}}].

\bibitem{2017MNRAS.466..581P}
J.~{Patterson}, A.~{Oksanen}, J.~{Kemp}, B.~{Monard}, R.~{Rea}, F.-J. {Hambsch}
  et~al., \emph{{T Pyxidis: death by a thousand novae}},
  \href{https://doi.org/10.1093/mnras/stw2970}{\emph{\mnras} {\bfseries 466}
  (2017) 581} [\href{https://arxiv.org/abs/1603.00291}{{\ttfamily
  1603.00291}}].

\bibitem{2018ApJ...862...89G}
P.~{Godon}, E.~M. {Sion}, R.~E. {Williams} and S.~{Starrfield}, \emph{{The
  Long-term Secular Mass Accretion Rate of the Recurrent Nova T Pyxidis}},
  \href{https://doi.org/10.3847/1538-4357/aacd0a}{\emph{\apj} {\bfseries 862}
  (2018) 89} [\href{https://arxiv.org/abs/1806.06059}{{\ttfamily 1806.06059}}].

\bibitem{1997A&A...323L..49P}
M.~A.~C. {Perryman}, L.~{Lindegren}, J.~{Kovalevsky}, E.~{Hoeg}, U.~{Bastian},
  P.~L. {Bernacca} et~al., \emph{{The HIPPARCOS Catalogue}}, {\emph{\aap}
  {\bfseries 323} (1997) L49}.

\bibitem{2006AJ....131.1163S}
M.~F. {Skrutskie}, R.~M. {Cutri}, R.~{Stiening}, M.~D. {Weinberg},
  S.~{Schneider}, J.~M. {Carpenter} et~al., \emph{{The Two Micron All Sky
  Survey (2MASS)}}, \href{https://doi.org/10.1086/498708}{\emph{\aj} {\bfseries
  131} (2006) 1163}.

\bibitem{2004ApJ...612..168P}
A.~{Pietrinferni}, S.~{Cassisi}, M.~{Salaris} and F.~{Castelli}, \emph{{A Large
  Stellar Evolution Database for Population Synthesis Studies. I. Scaled Solar
  Models and Isochrones}}, \href{https://doi.org/10.1086/422498}{\emph{\apj}
  {\bfseries 612} (2004) 168}
  [\href{https://arxiv.org/abs/astro-ph/0405193}{{\ttfamily
  astro-ph/0405193}}].

\bibitem{1929ApJ....69..103H}
E.~P. {Hubble}, \emph{{A spiral nebula as a stellar system, Messier 31.}},
  \href{https://doi.org/10.1086/143167}{\emph{\apj} {\bfseries 69} (1929) 103}.

\bibitem{1945PASP...57...69M}
D.~B. {Mclaughlin}, \emph{{The Relation between Light-Curves and Luminosities
  of Novae}}, \href{https://doi.org/10.1086/125689}{\emph{\pasp} {\bfseries 57}
  (1945) 69}.

\bibitem{2018A&A...616A...1G}
{Gaia Collaboration}, A.~G.~A. {Brown}, A.~{Vallenari}, T.~{Prusti}, J.~H.~J.
  {de Bruijne}, C.~{Babusiaux} et~al., \emph{{Gaia Data Release 2. Summary of
  the contents and survey properties}},
  \href{https://doi.org/10.1051/0004-6361/201833051}{\emph{\aap} {\bfseries
  616} (2018) A1} [\href{https://arxiv.org/abs/1804.09365}{{\ttfamily
  1804.09365}}].

\bibitem{2018MNRAS.481.3033S}
B.~E. {Schaefer}, \emph{{The distances to Novae as seen by Gaia}},
  \href{https://doi.org/10.1093/mnras/sty2388}{\emph{\mnras} {\bfseries 481}
  (2018) 3033} [\href{https://arxiv.org/abs/1809.00180}{{\ttfamily
  1809.00180}}].

\bibitem{2019A&A...622A.186S}
P.~{Selvelli} and R.~{Gilmozzi}, \emph{{A UV and optical study of 18 old novae
  with Gaia DR2 distances: mass accretion rates, physical parameters, and
  MMRD}}, \href{https://doi.org/10.1051/0004-6361/201834238}{\emph{\aap}
  {\bfseries 622} (2019) A186}
  [\href{https://arxiv.org/abs/1903.05868}{{\ttfamily 1903.05868}}].

\bibitem{2011ApJ...735...94K}
M.~M. {Kasliwal}, S.~B. {Cenko}, S.~R. {Kulkarni}, E.~O. {Ofek}, R.~{Quimby}
  and A.~{Rau}, \emph{{Discovery of a New Photometric Sub-class of Faint and
  Fast Classical Novae}},
  \href{https://doi.org/10.1088/0004-637X/735/2/94}{\emph{\apj} {\bfseries 735}
  (2011) 94} [\href{https://arxiv.org/abs/1003.1720}{{\ttfamily 1003.1720}}].

\bibitem{2017ApJ...839..109S}
M.~M. {Shara}, T.~{Doyle}, T.~R. {Lauer}, D.~{Zurek}, E.~A. {Baltz},
  A.~{Kovetz} et~al., \emph{{A Hubble Space Telescope Survey for Novae in M87.
  II. Snuffing out the Maximum Magnitude-Rate of Decline Relation for Novae as
  a Non-standard Candle, and a Prediction of the Existence of Ultrafast
  Novae}}, \href{https://doi.org/10.3847/1538-4357/aa65cd}{\emph{\apj}
  {\bfseries 839} (2017) 109}
  [\href{https://arxiv.org/abs/1702.05788}{{\ttfamily 1702.05788}}].

\bibitem{2017ApJ...839...88K}
M.~M. {Kasliwal}, J.~{Bally}, F.~{Masci}, A.~M. {Cody}, H.~E. {Bond}, J.~E.
  {Jencson} et~al., \emph{{SPIRITS: Uncovering Unusual Infrared Transients with
  Spitzer}}, \href{https://doi.org/10.3847/1538-4357/aa6978}{\emph{\apj}
  {\bfseries 839} (2017) 88}
  [\href{https://arxiv.org/abs/1701.01151}{{\ttfamily 1701.01151}}].

\bibitem{1973ApJ...186.1007W}
J.~{Whelan} and I.~{Iben}, Jr., \emph{{Binaries and Supernovae of Type I}},
  \href{https://doi.org/10.1086/152565}{\emph{\apj} {\bfseries 186} (1973)
  1007}.

\bibitem{1999ApJ...519..314H}
I.~{Hachisu}, M.~{Kato}, K.~{Nomoto} and H.~{Umeda}, \emph{{A New Evolutionary
  Path to Type IA Supernovae: A Helium-rich Supersoft X-Ray Source Channel}},
  \href{https://doi.org/10.1086/307370}{\emph{\apj} {\bfseries 519} (1999) 314}
  [\href{https://arxiv.org/abs/astro-ph/9902303}{{\ttfamily
  astro-ph/9902303}}].

\bibitem{1999ApJ...522..487H}
I.~{Hachisu}, M.~{Kato} and K.~{Nomoto}, \emph{{A Wide Symbiotic Channel to
  Type IA Supernovae}}, \href{https://doi.org/10.1086/307608}{\emph{\apj}
  {\bfseries 522} (1999) 487}
  [\href{https://arxiv.org/abs/astro-ph/9902304}{{\ttfamily
  astro-ph/9902304}}].

\bibitem{2000ARA&A..38..191H}
W.~{Hillebrandt} and J.~C. {Niemeyer}, \emph{{Type IA Supernova Explosion
  Models}}, \href{https://doi.org/10.1146/annurev.astro.38.1.191}{\emph{\araa}
  {\bfseries 38} (2000) 191}
  [\href{https://arxiv.org/abs/astro-ph/0006305}{{\ttfamily
  astro-ph/0006305}}].

\bibitem{2014ApJ...788..164P}
A.~{Pagnotta} and B.~E. {Schaefer}, \emph{{Identifying and Quantifying
  Recurrent Novae Masquerading as Classical Novae}},
  \href{https://doi.org/10.1088/0004-637X/788/2/164}{\emph{\apj} {\bfseries
  788} (2014) 164} [\href{https://arxiv.org/abs/1405.0246}{{\ttfamily
  1405.0246}}].

\bibitem{2014ApJS..213...10W}
S.~C. {Williams}, M.~J. {Darnley}, M.~F. {Bode}, A.~{Keen} and A.~W. {Shafter},
  \emph{{On the Progenitors of Local Group Novae. I. The M31 Catalog}},
  \href{https://doi.org/10.1088/0067-0049/213/1/10}{\emph{\apjs} {\bfseries
  213} (2014) 10} [\href{https://arxiv.org/abs/1405.4874}{{\ttfamily
  1405.4874}}].

\bibitem{2016ApJ...817..143W}
S.~C. {Williams}, M.~J. {Darnley}, M.~F. {Bode} and A.~W. {Shafter}, \emph{{On
  the Progenitors of Local Group Novae. II. The Red Giant Nova Rate of M31}},
  \href{https://doi.org/10.3847/0004-637X/817/2/143}{\emph{\apj} {\bfseries
  817} (2016) 143} [\href{https://arxiv.org/abs/1512.04088}{{\ttfamily
  1512.04088}}].

\bibitem{2018A&A...619A.121C}
J.~{Casanova}, J.~{Jos{\'e}} and S.~N. {Shore}, \emph{{Two-dimensional
  simulations of mixing in classical novae: The effect of white dwarf
  composition and mass}},
  \href{https://doi.org/10.1051/0004-6361/201833422}{\emph{\aap} {\bfseries
  619} (2018) A121} [\href{https://arxiv.org/abs/1807.10646}{{\ttfamily
  1807.10646}}].

\bibitem{1994ApJ...424..319K}
A.~{Kovetz} and D.~{Prialnik}, \emph{{Accretion onto a 1.4 solar mass white
  dwarf: Classical nova, recurrent nova, or supernova?}},
  \href{https://doi.org/10.1086/173891}{\emph{\apj} {\bfseries 424} (1994)
  319}.

\bibitem{1995ApJ...445..789P}
D.~{Prialnik} and A.~{Kovetz}, \emph{{An extended grid of multicycle nova
  evolution models}}, \href{https://doi.org/10.1086/175741}{\emph{\apj}
  {\bfseries 445} (1995) 789}.

\bibitem{2012BASI...40..419S}
S.~{Starrfield}, C.~{Iliadis}, F.~X. {Timmes}, W.~R. {Hix}, W.~D. {Arnett},
  C.~{Meakin} et~al., \emph{{Theoretical studies of accretion of matter onto
  white dwarfs and the single degenerate scenario for supernovae of Type Ia}},
  {\emph{Bulletin of the Astronomical Society of India} {\bfseries 40} (2012)
  419} [\href{https://arxiv.org/abs/1210.6086}{{\ttfamily 1210.6086}}].

\bibitem{2015MNRAS.446.1924H}
Y.~{Hillman}, D.~{Prialnik}, A.~{Kovetz} and M.~M. {Shara},
  \emph{{Observational signatures of SNIa progenitors, as predicted by
  models}}, \href{https://doi.org/10.1093/mnras/stu2235}{\emph{\mnras}
  {\bfseries 446} (2015) 1924}
  [\href{https://arxiv.org/abs/1411.0382}{{\ttfamily 1411.0382}}].

\bibitem{2015ApJ...808...52K}
M.~{Kato}, H.~{Saio} and I.~{Hachisu}, \emph{{Multi-wavelength Light Curve
  Model of the One-year Recurrence Period Nova M31N 2008-12A}},
  \href{https://doi.org/10.1088/0004-637X/808/1/52}{\emph{\apj} {\bfseries 808}
  (2015) 52} [\href{https://arxiv.org/abs/1506.05364}{{\ttfamily 1506.05364}}].

\bibitem{2017ApJ...844..143K}
M.~{Kato}, H.~{Saio} and I.~{Hachisu}, \emph{{A Millennium-long Evolution of
  the 1 yr Recurrence Period Nova{\textemdash}Search for Any Indication of the
  Forthcoming He Flash}},
  \href{https://doi.org/10.3847/1538-4357/aa7c5e}{\emph{\apj} {\bfseries 844}
  (2017) 143} [\href{https://arxiv.org/abs/1706.08654}{{\ttfamily
  1706.08654}}].

\bibitem{2013MNRAS.433.2884I}
I.~{Idan}, N.~J. {Shaviv} and G.~{Shaviv}, \emph{{The fate of a WD accreting
  H-rich material at high accretion rates}},
  \href{https://doi.org/10.1093/mnras/stt908}{\emph{\mnras} {\bfseries 433}
  (2013) 2884}.

\bibitem{2014ASPC..490..287N}
G.~{Newsham}, S.~{Starrfield} and F.~X. {Timmes}, \emph{{Evolution of Accreting
  White Dwarfs: Some of Them Continue to Grow}},  in \emph{Stellar Novae: Past
  and Future Decades}, P.~A. {Woudt} and V.~A.~R.~M. {Ribeiro}, eds., vol.~490
  of \emph{Astronomical Society of the Pacific Conference Series}, p.~287,
  Dec., 2014, \href{https://arxiv.org/abs/1303.3642}{{\ttfamily 1303.3642}}.

\bibitem{HILLMAN2019}
Y.~Hillman, M.~Shara, D.~Prialnik and A.~Kovetz, \emph{Multi-outburst nova
  modeling \& where models meet observations},
  \href{https://doi.org/https://doi.org/10.1016/j.asr.2019.08.029}{\emph{Advances
  in Space Research} (2019) }.

\bibitem{2010ApJS..187..275S}
B.~E. {Schaefer}, \emph{{Comprehensive Photometric Histories of All Known
  Galactic Recurrent Novae}},
  \href{https://doi.org/10.1088/0067-0049/187/2/275}{\emph{\apjs} {\bfseries
  187} (2010) 275} [\href{https://arxiv.org/abs/0912.4426}{{\ttfamily
  0912.4426}}].

\bibitem{2008ASPC..401...31A}
G.~C. {Anupama}, \emph{{The Recurrent Nova Class of Objects}},  in \emph{RS
  Ophiuchi (2006) and the Recurrent Nova Phenomenon}, A.~{Evans}, M.~F. {Bode},
  T.~J. {O'Brien} and M.~J. {Darnley}, eds., vol.~401 of \emph{Astronomical
  Society of the Pacific Conference Series}, pp.~31--41, 2008.

\bibitem{2010ApJ...724..480H}
R.~{Hounsell}, M.~F. {Bode}, P.~P. {Hick}, A.~{Buffington}, B.~V. {Jackson},
  J.~M. {Clover} et~al., \emph{{Exquisite Nova Light Curves from the Solar Mass
  Ejection Imager (SMEI)}},
  \href{https://doi.org/10.1088/0004-637X/724/1/480}{\emph{\apj} {\bfseries
  724} (2010) 480} [\href{https://arxiv.org/abs/1009.1737}{{\ttfamily
  1009.1737}}].

\bibitem{2012A&A...537A..34J}
R.~{Jurdana-{\v S}epi{\'c}}, V.~A.~R.~M. {Ribeiro}, M.~J. {Darnley},
  U.~{Munari} and M.~F. {Bode}, \emph{{Historical light curve and search for
  previous outbursts of Nova KT Eridani (2009)}},
  \href{https://doi.org/10.1051/0004-6361/201117806}{\emph{\aap} {\bfseries
  537} (2012) A34} [\href{https://arxiv.org/abs/1110.4637}{{\ttfamily
  1110.4637}}].

\bibitem{2013MNRAS.433.1991R}
V.~A.~R.~M. {Ribeiro}, M.~F. {Bode}, M.~J. {Darnley}, R.~M. {Barnsley},
  U.~{Munari} and D.~J. {Harman}, \emph{{Morpho-kinematical modelling of Nova
  Eridani 2009 (KT Eri)}},
  \href{https://doi.org/10.1093/mnras/stt856}{\emph{\mnras} {\bfseries 433}
  (2013) 1991} [\href{https://arxiv.org/abs/1305.3834}{{\ttfamily 1305.3834}}].

\bibitem{2014A&A...564A..76M}
U.~{Munari}, E.~{Mason} and P.~{Valisa}, \emph{{The narrow and moving HeII
  lines in nova KT Eridani}},
  \href{https://doi.org/10.1051/0004-6361/201323180}{\emph{\aap} {\bfseries
  564} (2014) A76} [\href{https://arxiv.org/abs/1403.3284}{{\ttfamily
  1403.3284}}].

\bibitem{2010MNRAS.401..121P}
K.~L. {Page}, J.~P. {Osborne}, P.~A. {Evans}, G.~A. {Wynn}, A.~P. {Beardmore},
  R.~L.~C. {Starling} et~al., \emph{{Swift observations of the X-ray and UV
  evolution of V2491 Cyg (Nova Cyg 2008 No. 2)}},
  \href{https://doi.org/10.1111/j.1365-2966.2009.15681.x}{\emph{\mnras}
  {\bfseries 401} (2010) 121}
  [\href{https://arxiv.org/abs/0909.1501}{{\ttfamily 0909.1501}}].

\bibitem{2011MNRAS.412.1701R}
V.~A.~R.~M. {Ribeiro}, M.~J. {Darnley}, M.~F. {Bode}, U.~{Munari}, D.~J.
  {Harman}, I.~A. {Steele} et~al., \emph{{The morphology of the expanding
  ejecta of V2491 Cygni (2008 N.2)}},
  \href{https://doi.org/10.1111/j.1365-2966.2010.18006.x}{\emph{\mnras}
  {\bfseries 412} (2011) 1701}
  [\href{https://arxiv.org/abs/1011.2045}{{\ttfamily 1011.2045}}].

\bibitem{2011NewA...16..209M}
U.~{Munari}, A.~{Siviero}, S.~{Dallaporta}, G.~{Cherini}, P.~{Valisa} and
  L.~{Tomasella}, \emph{{An extensive optical study of V2491 Cyg (Nova Cyg 2008
  N.2), from maximum brightness to return to quiescence}},
  \href{https://doi.org/10.1016/j.newast.2010.08.010}{\emph{\na} {\bfseries 16}
  (2011) 209} [\href{https://arxiv.org/abs/1009.0822}{{\ttfamily 1009.0822}}].

\bibitem{2011A&A...530A..70D}
M.~J. {Darnley}, V.~A.~R.~M. {Ribeiro}, M.~F. {Bode} and U.~{Munari}, \emph{{On
  the progenitor system of Nova V2491 Cygni}},
  \href{https://doi.org/10.1051/0004-6361/201016038}{\emph{\aap} {\bfseries
  530} (2011) A70} [\href{https://arxiv.org/abs/1104.3482}{{\ttfamily
  1104.3482}}].

\bibitem{2008ASPC..401..206H}
I.~{Hachisu}, M.~{Kato}, S.~{Kiyota}, K.~{Kubotera}, H.~{Maehara},
  K.~{Nakajima} et~al., \emph{{Optical Light Curves of RS Oph (2006) and
  Hydrogen Burning Turnoff}},  in \emph{RS Ophiuchi (2006) and the Recurrent
  Nova Phenomenon}, A.~{Evans}, M.~F. {Bode}, T.~J. {O'Brien} and M.~J.
  {Darnley}, eds., vol.~401 of \emph{Astronomical Society of the Pacific
  Conference Series}, (San Francisco), pp.~206--209, Dec., 2008,
  \href{https://arxiv.org/abs/0807.1240}{{\ttfamily 0807.1240}}.

\bibitem{2011MNRAS.414.2195A}
S.~{Adamakis}, S.~P.~S. {Eyres}, A.~{Sarkar} and R.~W. {Walsh}, \emph{{A
  pre-outburst signal in the long-term optical light curve of the recurrent
  nova RS Ophiuchi}},
  \href{https://doi.org/10.1111/j.1365-2966.2011.18536.x}{\emph{\mnras}
  {\bfseries 414} (2011) 2195}
  [\href{https://arxiv.org/abs/1102.3779}{{\ttfamily 1102.3779}}].

\bibitem{2010AN....331..187P}
W.~{Pietsch}, \emph{{X-ray emission from optical novae in M 31}},
  \href{https://doi.org/10.1002/asna.200911324}{\emph{Astronomische
  Nachrichten} {\bfseries 331} (2010) 187}
  [\href{https://arxiv.org/abs/0910.3865}{{\ttfamily 0910.3865}}].

\bibitem{1985ApJ...295..287D}
G.~{de Vaucouleurs} and H.~G. {Corwin}, Jr., \emph{{S Andromedae 1885 - A
  centennial review}}, \href{https://doi.org/10.1086/163374}{\emph{\apj}
  {\bfseries 295} (1985) 287}.

\bibitem{2015ApJ...814....3D}
J.~J. {Dalcanton}, M.~{Fouesneau}, D.~W. {Hogg}, D.~{Lang}, A.~K. {Leroy},
  K.~D. {Gordon} et~al., \emph{{The Panchromatic Hubble Andromeda Treasury.
  VIII. A Wide-area, High-resolution Map of Dust Extinction in M31}},
  \href{https://doi.org/10.1088/0004-637X/814/1/3}{\emph{\apj} {\bfseries 814}
  (2015) 3} [\href{https://arxiv.org/abs/1509.06988}{{\ttfamily 1509.06988}}].

\bibitem{tramp}
M.~J. {Darnley}, A.~M. {Newsam}, K.~{Chinetti}, I.~D.~W. {Hawkins}, A.~L.
  {Jannetta}, M.~M. {Kasliwal} et~al., \emph{{AT}\,2016dah and {AT}\,2017fyp:\
  the first classical novae discovered within in tidal stream},
  {\emph{Submitted, for publication in MNRAS} (2020) }.

\bibitem{2001Natur.412...49I}
R.~{Ibata}, M.~{Irwin}, G.~{Lewis}, A.~M.~N. {Ferguson} and N.~{Tanvir},
  \emph{{A giant stream of metal-rich stars in the halo of the galaxy M31}},
  {\emph{\nat} {\bfseries 412} (2001) 49}
  [\href{https://arxiv.org/abs/astro-ph/0107090}{{\ttfamily
  astro-ph/0107090}}].

\bibitem{2011ApJ...734...12S}
A.~W. {Shafter}, M.~J. {Darnley}, K.~{Hornoch}, A.~V. {Filippenko}, M.~F.
  {Bode}, R.~{Ciardullo} et~al., \emph{{A Spectroscopic and Photometric Survey
  of Novae in M31}},
  \href{https://doi.org/10.1088/0004-637X/734/1/12}{\emph{\apj} {\bfseries 734}
  (2011) 12} [\href{https://arxiv.org/abs/1104.0222}{{\ttfamily 1104.0222}}].

\bibitem{Ransome2019}
C.~{Ransome}, M.~J. {Darnley}, S.~M. {Habergham-Mawson}, M.~W. {Healy}, P.~A.
  {James}, S.~C. {Williams} et~al., ``{A Spectroscopic and Photometric Survey
  of Novae in M31 II}.'' 2020.

\bibitem{2015ATel.7116....1H}
K.~{Hornoch} and A.~W. {Shafter}, \emph{{M31N 2006-11c appears to be spatially
  coincident with PNV J00413317+4110124 and hence a recurrent nova in M31}},
  {\emph{The Astronomer's Telegram} {\bfseries 7116} (2015) }.

\bibitem{Sch2017}
P.~{Schmeer}, \emph{{CBAT}},  Dec., 2017.

\bibitem{2017ATel11088....1W}
S.~C. {Williams} and M.~J. {Darnley}, \emph{{Spectroscopic classification AT
  2017jdm as a nova, and likely recurrent eruption of M31N 2007-10b}},
  {\emph{The Astronomer's Telegram} {\bfseries 11088} (2017) }.

\bibitem{2017ATel10001....1S}
P.~{Sin}, M.~{Henze}, G.~{Sala}, A.~{Ederoclite}, M.~{Hernanz}, J.~{Jose}
  et~al., \emph{{Additional Photometry for nova M31N 2016-12e and
  classification as a recurrent nova (= M31N 2007-11f)}}, {\emph{ATel,
  No.~10001} {\bfseries 1} (2017) }.

\bibitem{2017RNAAS...1a..44S}
A.~W. {Shafter}, M.~{Henze}, M.~J. {Darnley}, R.~{Ciardullo}, B.~D. {Davis} and
  S.~L. {Hawley}, \emph{{The Recurrent Nova Candidate M31N 1966-08a = 1968-10c
  is a Galactic Flare Star}},
  \href{https://doi.org/10.3847/2515-5172/aaa086}{\emph{Research Notes of the
  American Astronomical Society} {\bfseries 1} (2017) 44}
  [\href{https://arxiv.org/abs/1712.05023}{{\ttfamily 1712.05023}}].

\bibitem{2012ApJ...752..156S}
A.~W. {Shafter}, M.~J. {Darnley}, M.~F. {Bode} and R.~{Ciardullo}, \emph{{On
  the Spectroscopic Classes of Novae in M33}},
  \href{https://doi.org/10.1088/0004-637X/752/2/156}{\emph{\apj} {\bfseries
  752} (2012) 156} [\href{https://arxiv.org/abs/1204.4850}{{\ttfamily
  1204.4850}}].

\bibitem{2016ApJS..222....9M}
P.~{Mr{\'o}z}, A.~{Udalski}, R.~{Poleski}, I.~{Soszy{\'n}ski}, M.~K.
  {Szyma{\'n}ski}, G.~{Pietrzy{\'n}ski} et~al., \emph{{OGLE Atlas of Classical
  Novae. II. Magellanic Clouds}},
  \href{https://doi.org/10.3847/0067-0049/222/1/9}{\emph{\apjs} {\bfseries 222}
  (2016) 9} [\href{https://arxiv.org/abs/1511.06355}{{\ttfamily 1511.06355}}].

\bibitem{2004AJ....127..816N}
J.~D. {Neill} and M.~M. {Shara}, \emph{{The H{$\alpha$} Light Curves and
  Spatial Distribution of Novae in M81}},
  \href{https://doi.org/10.1086/381484}{\emph{\aj} {\bfseries 127} (2004) 816}
  [\href{https://arxiv.org/abs/astro-ph/0311327}{{\ttfamily
  astro-ph/0311327}}].

\bibitem{2009ApJ...705.1056B}
M.~F. {Bode}, M.~J. {Darnley}, A.~W. {Shafter}, K.~L. {Page}, O.~{Smirnova},
  G.~C. {Anupama} et~al., \emph{{Optical and X-ray Observations of M31N
  2007-12b: An Extragalactic Recurrent Nova with a Detected Progenitor?}},
  \href{https://doi.org/10.1088/0004-637X/705/1/1056}{\emph{\apj} {\bfseries
  705} (2009) 1056} [\href{https://arxiv.org/abs/0902.0301}{{\ttfamily
  0902.0301}}].

\bibitem{2014ASPC..490...49D}
M.~J. {Darnley}, M.~F. {Bode}, D.~J. {Harman}, R.~A. {Hounsell}, U.~{Munari},
  V.~A.~R.~M. {Ribeiro} et~al., \emph{{On the Galactic Nova Progenitor
  Population}},  in \emph{Stellar Novae: Past and Future Decades}, P.~A.
  {Woudt} and V.~A.~R.~M. {Ribeiro}, eds., vol.~490 of \emph{Astronomical
  Society of the Pacific Conference Series}, (San Francisco), pp.~49--55, Dec.,
  2014, \href{https://arxiv.org/abs/1303.2711}{{\ttfamily 1303.2711}}.

\bibitem{2019MNRAS.tmp.2562K}
N.~P.~M. {Kuin}, K.~L. {Page}, P.~{Mr{\'o}z}, M.~J. {Darnley}, S.~N. {Shore},
  J.~P. {Osborne} et~al., \emph{{The January 2016 eruption of recurrent nova
  LMC 1968}}, \href{https://doi.org/10.1093/mnras/stz2960}{\emph{\mnras} (2019)
  2562} [\href{https://arxiv.org/abs/1909.03281}{{\ttfamily 1909.03281}}].

\bibitem{1995cvs..book.....W}
B.~{Warner}, \emph{{Cataclysmic variable stars}}. Cambridge Astrophysics
  Series, Cambridge, New York: Cambridge University Press, 1995, 1995.

\bibitem{2004SPIE.5489..679S}
I.~A. {Steele}, R.~J. {Smith}, P.~C. {Rees}, I.~P. {Baker}, S.~D. {Bates},
  M.~F. {Bode} et~al., \emph{{The Liverpool Telescope: performance and first
  results}},  in \emph{Ground-based Telescopes}, J.~M. {Oschmann}, Jr., ed.,
  vol.~5489 of \emph{Society of Photo-Optical Instrumentation Engineers (SPIE)
  Conference Series}, pp.~679--692, Oct., 2004,
  \href{https://doi.org/10.1117/12.551456}{DOI}.

\bibitem{2004ApJ...611.1005G}
N.~{Gehrels}, G.~{Chincarini}, P.~{Giommi}, K.~O. {Mason}, J.~A. {Nousek},
  A.~A. {Wells} et~al., \emph{{The Swift Gamma-Ray Burst Mission}},
  \href{https://doi.org/10.1086/422091}{\emph{\apj} {\bfseries 611} (2004)
  1005}.

\bibitem{PAGE2019}
K.~Page, A.~Beardmore and J.~Osborne, \emph{Neil gehrels swift observatory
  studies of supersoft novae},
  \href{https://doi.org/https://doi.org/10.1016/j.asr.2019.08.003}{\emph{Advances
  in Space Research} (2019) }.

\bibitem{2012arXiv1209.3114M}
A.~{Merloni}, P.~{Predehl}, W.~{Becker}, H.~{B{\"o}hringer}, T.~{Boller},
  H.~{Brunner} et~al., \emph{{eROSITA Science Book: Mapping the Structure of
  the Energetic Universe}}, {\emph{arXiv e-prints} (2012) arXiv:1209.3114}
  [\href{https://arxiv.org/abs/1209.3114}{{\ttfamily 1209.3114}}].

\bibitem{2019ApJ...873..111I}
{\v{Z}}.~{Ivezi{\'c}}, S.~M. {Kahn}, J.~A. {Tyson}, B.~{Abel}, E.~{Acosta},
  R.~{Allsman} et~al., \emph{{LSST: From Science Drivers to Reference Design
  and Anticipated Data Products}},
  \href{https://doi.org/10.3847/1538-4357/ab042c}{\emph{\apj} {\bfseries 873}
  (2019) 111} [\href{https://arxiv.org/abs/0805.2366}{{\ttfamily 0805.2366}}].

\bibitem{2015ExA....39..119C}
C.~M. {Copperwheat}, I.~A. {Steele}, R.~M. {Barnsley}, S.~D. {Bates},
  D.~{Bersier}, M.~F. {Bode} et~al., \emph{{Liverpool telescope 2: a new
  robotic facility for rapid transient follow-up}},
  \href{https://doi.org/10.1007/s10686-015-9447-0}{\emph{Experimental
  Astronomy} {\bfseries 39} (2015) 119}
  [\href{https://arxiv.org/abs/1410.1731}{{\ttfamily 1410.1731}}].

\bibitem{BSG}
R.~D. {Moore}, D.~{Eick}, H.~{Frand}, R.~E. {French} and M.~{Rymer},
  \emph{Battlestar Galactica:\ Season 4 --- No Exit}. NBCUniversal Television
  Distribution, 2009.

\bibitem{2008Nis}
K.~{Nishiyama} and F.~{Kabashima}, \emph{{CBAT}},  Dec., 2008.

\bibitem{2019ATel13269....1O}
A.~{Oksanen}, M.~J. {Darnley}, A.~W. {Shafter}, S.~{Kafka}, M.~{Kato} and
  M.~{Henze}, \emph{{Recurrent Nova M31N 2008-12a: discovery of the 2019
  eruption}}, {\emph{The Astronomer's Telegram} {\bfseries 13269} (2019) 1}.

\bibitem{2014A&A...563L...8H}
M.~{Henze}, J.-U. {Ness}, M.~J. {Darnley}, M.~F. {Bode}, S.~C. {Williams},
  A.~W. {Shafter} et~al., \emph{{A remarkable recurrent nova in M 31: The X-ray
  observations}},
  \href{https://doi.org/10.1051/0004-6361/201423410}{\emph{\aap} {\bfseries
  563} (2014) L8} [\href{https://arxiv.org/abs/1401.2904}{{\ttfamily
  1401.2904}}].

\bibitem{2014ApJ...786...61T}
S.~{Tang}, L.~{Bildsten}, W.~M. {Wolf}, K.~L. {Li}, A.~K.~H. {Kong}, Y.~{Cao}
  et~al., \emph{{An Accreting White Dwarf near the Chandrasekhar Limit in the
  Andromeda Galaxy}},
  \href{https://doi.org/10.1088/0004-637X/786/1/61}{\emph{\apj} {\bfseries 786}
  (2014) 61} [\href{https://arxiv.org/abs/1401.2426}{{\ttfamily 1401.2426}}].

\bibitem{2017ApJ...849...96D}
M.~J. {Darnley}, R.~{Hounsell}, P.~{Godon}, D.~A. {Perley}, M.~{Henze},
  N.~P.~M. {Kuin} et~al., \emph{{Inflows, Outflows, and a Giant Donor in the
  Remarkable Recurrent Nova M31N 2008-12a?---Hubble Space Telescope Photometry
  of the 2015 Eruption}},
  \href{https://doi.org/10.3847/1538-4357/aa9062}{\emph{\apj} {\bfseries 849}
  (2017) 96} [\href{https://arxiv.org/abs/1709.10145}{{\ttfamily 1709.10145}}].

\bibitem{1985MNRAS.212..753W}
R.~E. {Williams}, E.~P. {Ney}, W.~M. {Sparks}, S.~G. {Starrfield}, S.~{Wyckoff}
  and J.~W. {Truran}, \emph{{Ultraviolet spectral evolution and heavy element
  abundances in nova Coronae Austrinae 1981.}},
  \href{https://doi.org/10.1093/mnras/212.4.753}{\emph{\mnras} {\bfseries 212}
  (1985) 753}.

\bibitem{1986ApJ...303L...5S}
S.~{Starrfield}, W.~M. {Sparks} and J.~W. {Truran}, \emph{{Hydrodynamic Models
  for Novae with Ejecta Rich in Oxygen, Neon, and Magnesium}},
  \href{https://doi.org/10.1086/184642}{\emph{\apjl} {\bfseries 303} (1986)
  L5}.

\bibitem{2017ApJ...847...35D}
M.~J. {Darnley}, R.~{Hounsell}, P.~{Godon}, D.~A. {Perley}, M.~{Henze},
  N.~P.~M. {Kuin} et~al., \emph{{No Neon, but Jets in the Remarkable Recurrent
  Nova M31N 2008-12a?---Hubble Space Telescope Spectroscopy of the 2015
  Eruption}}, \href{https://doi.org/10.3847/1538-4357/aa8867}{\emph{\apj}
  {\bfseries 847} (2017) 35}
  [\href{https://arxiv.org/abs/1708.06795}{{\ttfamily 1708.06795}}].

\bibitem{2016ApJ...833..149D}
M.~J. {Darnley}, M.~{Henze}, M.~F. {Bode}, I.~{Hachisu}, M.~{Hernanz},
  K.~{Hornoch} et~al., \emph{{M31N 2008-12a - The Remarkable Recurrent Nova in
  M31: Panchromatic Observations of the 2015 Eruption.}},
  \href{https://doi.org/10.3847/1538-4357/833/2/149}{\emph{\apj} {\bfseries
  833} (2016) 149} [\href{https://arxiv.org/abs/1607.08082}{{\ttfamily
  1607.08082}}].

\bibitem{2017ASPC..509..515D}
M.~J. {Darnley}, \emph{{M31N 2008-12a --- The Remarkable Recurrent Nova in
  M31}},  in \emph{20th European White Dwarf Workshop}, P.-E. {Tremblay},
  B.~{Gaensicke} and T.~{Marsh}, eds., vol.~509 of \emph{Astronomical Society
  of the Pacific Conference Series}, (San Francisco), pp.~515--520, Mar., 2017,
  \href{https://arxiv.org/abs/1611.01301}{{\ttfamily 1611.01301}}.

\bibitem{12a1718}
M.~J. {Darnley}, M.~{Henze}, A.~W. {Shafter}, S.~C. {Williams}, D.~{Boyd} and
  K.~{Hornoch}, ``{Back on Track? --- The 2017 and 2018 eruptions of M31N
  2008-12a}.'' 2020.

\bibitem{2015A&A...580A..45D}
M.~J. {Darnley}, M.~{Henze}, I.~A. {Steele}, M.~F. {Bode}, V.~A.~R.~M.
  {Ribeiro}, P.~{Rodr{\'{\i}}guez-Gil} et~al., \emph{{A remarkable recurrent
  nova in M31: Discovery and optical/UV observations of the predicted 2014
  eruption}}, \href{https://doi.org/10.1051/0004-6361/201526027}{\emph{\aap}
  {\bfseries 580} (2015) A45}
  [\href{https://arxiv.org/abs/1506.04202}{{\ttfamily 1506.04202}}].

\bibitem{2019Natur.565..460D}
M.~J. {Darnley}, R.~{Hounsell}, T.~J. {O'Brien}, M.~{Henze},
  P.~{Rodr{\'{\i}}guez-Gil}, A.~W. {Shafter} et~al., \emph{{A recurrent nova
  super-remnant in the Andromeda galaxy}},
  \href{https://doi.org/10.1038/s41586-018-0825-4}{\emph{\nat} {\bfseries 565}
  (2019) 460} [\href{https://arxiv.org/abs/1712.04872}{{\ttfamily
  1712.04872}}].

\end{thebibliography}\endgroup

\end{document}